\documentclass[review]{elsarticle}
\graphicspath{ {./figures/} }
\usepackage{hyperref}
\usepackage{float}
\usepackage{verbatim} 
\usepackage{apalike}
\restylefloat{figure}
\restylefloat{table}

\usepackage{amsmath,amssymb,amsfonts}
\usepackage{subfigure}
\usepackage{multirow}
\usepackage{makecell}

\journal{Expert Systems with Applications}

\biboptions{authoryear}

\begin{document}
\begin{frontmatter}

\begin{titlepage}
\begin{center}
\vspace*{1cm}


\textbf{Robust image steganography against lossy JPEG compression based on embedding domain selection and adaptive error correction}

\vspace{1.5cm}

Xiaolong Duan$^{a}$ (xiaolongduan@foxmail.com), Bin Li$^b$ (libin@szu.edu.cn), Zhaoxia Yin$^c$ (zxyin@cee.ecnu.edu.cn), Xinpeng Zhang$^d$ (zhangxinpeng@fudan.edu.cn), Bin Luo$^a$ (luobin@ahu.edu.cn) \\

\hspace{10pt}

\begin{flushleft}
\small  
$^a$ Anhui Province Key Laboratory of Multimodal Cognitive Computation, School of Computer Science and Technology, Anhui University, Hefei, 230601, China \\
$^b$ Guangdong Key Laboratory of Intelligent Information Processing and Shenzhen Key Laboratory of Media Security, Shenzhen University, Shenzhen, 518060, China \\
$^c$  School of Communication \& Electronic Engineering, East China Normal University, Shanghai, 200241, China \\
$^d$  School of Computer Science, Fudan University, Shanghai, 200433, China

\vspace{1cm}
\textbf{Corresponding Author:} \\
Zhaoxia Yin \\
School of Communication \& Electronic Engineering, East China Normal University, Shanghai, 200241, China \\
Tel:  \\
Email: zxyin@cee.ecnu.edu.cn

\end{flushleft}        
\end{center}
\end{titlepage}

\title{Robust image steganography against lossy JPEG compression based on embedding domain selection and adaptive error correction}

\author[label1]{Xiaolong Duan}
\ead{xiaolongduan@foxmail.com}

\author[label2]{Bin Li}
\ead{libin@szu.edu.cn}

\author[label3]{Zhaoxia Yin \corref{cor1}}
\ead{zxyin@cee.ecnu.edu.cn}
\cortext[cor1]{Corresponding author.}

\author[label4]{Xinpeng Zhang}
\ead{zhangxinpeng@fudan.edu.cn}

\author[label1]{Bin Luo}
\ead{luobin@ahu.edu.cn}

\address[label1]{Anhui Province Key Laboratory of Multimodal Cognitive Computation, School of Computer Science and Technology, Anhui University, Hefei, 230601, China}
\address[label2]{Guangdong Key Laboratory of Intelligent Information Processing and Shenzhen Key Laboratory of Media Security, Shenzhen University, Shenzhen, 518060, China}
\address[label3]{School of Communication \& Electronic Engineering, East China Normal University, Shanghai, 200241, China}
\address[label4]{School of Computer Science, Fudan University, Shanghai, 200433, China}

\begin{abstract}
Transmitting images for communication on social networks has become routine, which is helpful for covert communication. The traditional steganography algorithm is unable to successfully convey secret information since the social network channel will perform lossy operations on images, such as JPEG compression.
Previous studies tried to solve this problem by enhancing the robustness or making the cover adapt to the channel processing.
In this study, we proposed a robust image steganography method against lossy JPEG compression based on embedding domain selection and adaptive error correction.
To improve anti-steganalysis performance, the embedding domain is selected adaptively.
To increase robustness and lessen the impact on anti-steganalysis performance, the error correction capacity of the error correction code is adaptively adjusted to eliminate redundancy.
The experimental results show that the proposed method achieves better anti-steganalysis and robustness.

\end{abstract}

\begin{keyword}
Robust steganography \sep Embedding domain selection \sep Adaptive error correction \sep Dither modulation \sep Social networks
\end{keyword}

\end{frontmatter}


\section{Introduction}
\label{sec::introduction}
As one of the practice applications of data hiding technique \citep*{fanPixelTypeClassification2022,liuInvisibleRobustWatermarking2022,wangHighcapacityReversibleData2023}, steganography \citep*{liaoNewPayloadPartition2020a,tangCNNBasedAdversarialEmbedding2019,al-dmourSteganographyEmbeddingMethod2016,sImperceptibilityRobustnessTradeoff2016,benkhaddraSecureTransmissionSecret2023} algorithms embed secret information into digital multimedia and make it difficult to be detected while being transmitted over a channel. To detect steganography, steganalysis \citep*{holubLowComplexityFeaturesJPEG2015,geethaAudioSteganalysisHausdorff2010} is proposed, designing classification models to assess whether an image is clean or steganographic.  For steganographic images containing secret data embedded by a steganography algorithm, the higher the error judgment rate of a steganalysis  classification model,  the better the security of the steganography algorithm. So steganalysis test is also used to analyze the security  (imperceptibility and anti-steganalysis performance) of a steganography algorithm. 

In recent years, adaptive steganography algorithms have been created to resist steganalysis attacks \citep*{holubUniversalDistortionFunction2014,suNewDistortionFunction2018,fengDiversityBasedCascadeFilters2020}. The distortion function calculates the loss caused by employing steganography codes \citep*{fillerMinimizingAdditiveDistortion2011a} to integrate hidden information and convert an ordinary image into a steganographic image. The anti-steganalysis performance of steganography has been improved. However, to ensure that the secret information is retrieved accurately, most steganography techniques assume that the transmission channel is lossless.

Nowadays, a lot of data is transferred in the network; however, the data is likely to get attacked throughout the network channel transmission process, including by added noise and data compression.
The original signal is lost when digital data is compressed, such as JPEG compression, and the receiver cannot retrieve the intended cover.
As a result, if the cover signal is attacked, the receiver will be unable to decode the secret message precisely.
Some algorithms to destroy secret communication \citep*{wangDefeatingDataHiding2020a,zhuDestroyingRobustSteganography2021a} have also been proposed, which is a challenge to traditional steganography.
Many robust steganography methods \citep*{tsaiJointRobustnessSecurity2012,jen-shengtsaiSelectionOptimalFeature2011a,liuCoverlessSteganographyBased2020,liuAdversarialRobustImage2022} for JPEG image compression have been proposed to ensure that the information is accurately extracted by the receiver and that signal loss is avoided while the message transmits across the lossy channel. 
Robustness and anti-steganalysis performance are both important for steganography in social network channels.

Currently, multiple types of research are attempting to achieve robust adaptive steganography. They increase the robustness of the steganography algorithm by creating a robust cover and adapting it to the channel. ``Cover robustness" and ``Adaptive channel robustness" are discussed in this paper.

``Adaptive channel robustness" means that the processed image can adapt to the lossy operation of the channel, to improve the steganography robustness. In \citep*{zhaoImprovingRobustnessAdaptive2019}, Zhao {\textit{et al.}} proposed a transport channel matching based on robust steganography. They use the channel quality factor to re-compress the cover image numerous times before encoding a secret message. They use BCH (Bose, Ray-Chaudhuri, and Hocquenghem) code \citep*{forneyDecodingBCHCodes1965} to encrypt the secret message to increase the extraction accuracy of the secret message. The scheme is robust and secure. In addition, the method has a large capacity. However, it is a time-consuming activity and suspicious behavior. 
Tao {\textit{et al.}} proposed a method to solve the problem that the JPEG compression factor of the channel is greater than that of the cover in \citep*{taoRobustImageSteganography2019}.
They first compress the original image by JPEG with channel quality factor, and then use a steganography algorithm to embed information on the compressed image to produce a stego image.
According to the stego image, they adjust the DCT coefficient of the original image to form an intermediate image. Their goal is that the intermediate image is precisely the stego image generated after channel compression.
After channel JPEG compression, the method can completely extract the secret message and achieve excellent anti-steganalysis performance. However, the DCT coefficient of intermediate and original image residuals is too big to ensure anti-steganalysis performance.
In \citep*{luSecureRobustJPEG2020}, Lu {\textit{et al.}} trained an autoencoder model to learn the inverse process of JPEG compression. Using this model, the intermediate image can be generated, which is a stego image after channel compression.

The term ``Cover robustness" refers to the process of creating a cover with anti-JPEG compression and defining associated embedding rules to achieve robust steganography.
Zhang {\textit{et al.}} introduced the ``Compression-resistant Domain Constructing + RS-STC Codes" framework in \citep*{zhangFrameworkAdaptiveSteganography2016}, which provides robustness by using RS (Reed-Solomon) codes \citep*{macwilliams1977theory} and building robust embedding domains. They also investigated the possibility of embedding messages inside the framework utilizing the relative connection between four DCT coefficients in \citep*{zhangJPEGCompressionResistantAdaptive2015}.
In DMAS(Dither Modulation Based Robust Adaptive Steganography) \citep*{zhangDitherModulationBased2018}, according to quantization tables, Zhang {\textit{et al.}} operated dither modulation to adjust the middle-frequency AC coefficients using the features of the quantization process.
Based on DMAS, Yu {\textit{et al.}} introduced GMAS(Generalized dither Modulation based robust Adaptive Steganography) \citep*{yuRobustAdaptiveSteganography2020}. By substituting symmetric distortion with asymmetric distortion, they widen the embedding domain to lower frequency domains.
Zhang {\textit{et al.}} combine the framework with DMAS \citep*{zhangDitherModulationBased2018} to develop E-DMAS(Enhancing Dither Modulation based robust Adaptive Steganography) \citep*{zhangEnhancingReliabilityEfficiency2020}. E-DMAS can address the issue of anti-steganalysis performance rapidly degrading due to the usage of an increased number of check codes. 
Yin {\textit{et al.}} proposed MR-DMAS(Robust Adaptive Steganography Based on Dither Modulation and Modification with Re-compression) \citep*{yinRobustAdaptiveSteganography2021}, which is based on E-DMAS. They shifted the embedding domain to a lower frequency domain and added more check codes to boost robustness to balance robustness and anti-steganalysis performance.
SS(Sign Steganography) \citep*{zhuRobustSteganographyModifying2019}, ESS(Enhanced Sign Steganography) \citep*{qiaoRobustSteganographyResisting2021}, SSR(Sign Steganography Revisited) \citep*{wuSignSteganographyRevisited2022} use DCT coefficient symbols as robust cover elements, because coefficient symbols are robust in the compression process.


\begin{table*}[!t]
\caption{Comparison of two categories of robust steganography methods.}
\label{table_performance_three_schemes}
\centering
\resizebox{\textwidth}{!}{
\begin{tabular}{c c c c c c c c}
\hline
\bfseries Method & \bfseries Reference & \bfseries Anti-Steganalysis & \bfseries Robustness & \bfseries Capacity & \bfseries Security Flaw  & \bfseries    Reverse Engineering \\
\hline
Cover Robust & 
\makecell[c]{\citep*{zhangDitherModulationBased2018} \\ 
\citep*{yuRobustAdaptiveSteganography2020}
\\ \citep*{yinRobustAdaptiveSteganography2021}
}
& $\star$ & $\star\star\star$ & $\star$ & $\times$ & $\times$ \\
Adaptive Channel Robust & 

\makecell[c]{\citep*{zhaoImprovingRobustnessAdaptive2019} \\ 
\citep*{taoRobustImageSteganography2019}
\\ \citep*{luSecureRobustJPEG2020}
}

 & $\star\star\star$ & $\star\star$ & $\star\star\star$ & $\surd$ & $\surd$ \\
\hline
\end{tabular}
}
\end{table*}

The above two categories of methods have their own advantages and disadvantages, as shown in Table \ref{table_performance_three_schemes}.
``Adaptive channel robustness" has a security flaw that allows an attacker to build targeted steganalysis.
They must also assume that the JPEG encoder of the social network can be completely reverse engineered, which is called reverse engineering.
``Cover robustness" is now a scheme with routine behavior, low computational complexity, and high robustness that can be used in social networks without the issues outlined above. 
However, ``Cover robustness" has a small embedding capacity and weak anti-steganalysis performance.

In this paper, we improve the GMAS \citep*{yuRobustAdaptiveSteganography2020} method to improve the robustness and anti-steganalysis performance of steganography.
The embedding domain of the GMAS is fixed in the middle frequency domain, and the error correction capability is consistent across all images, resulting in weak anti-steganalysis performance. 
Because various images are damaged differently by JPEG compression and have varying robustness, fixed embedding domain, and error correction capabilities are unable to make appropriate use of the robustness of the image.
The fixed embedding domain in the middle frequency region can not make good use of the robustness of the image itself.
Fixed error correction capability causes error correction coding redundancy and affect steganography security for images with strong robustness. For images with weak robustness, the effect of the error correction code is not obvious.
We try to improve its embedding domain selection and error correction capacity, according to the robustness of the image.
Extensive experiments with CCPEV \citep*{kodovskyCalibrationRevisited2009} and DCTR \citep*{holubLowComplexityFeaturesJPEG2015}, as well as effective steganalysis, are used to assess the performance under various channel compression settings. According to the findings of the experiments, the proposed method outperforms GMAS in terms of robustness and anti-steganalysis performance.
The main contributions of this study are summarized as follows:
\begin{enumerate}[1)]
\item
To improve anti-steganalysis performance, the embedding domain of the dither modulation algorithm preferentially selects the region where the low-frequency DCT coefficient is located.
Because most images have strong robustness against JPEG compression, using a low frequency embedding domain is enough to ensure robustness, and the anti-steganalysis performance of the low frequency embedding domain is high.

\item
To improve the robustness and reduce the impact on anti-steganalysis performance, the proposed method adaptively adjusts the error correction capability according to the robustness of the image against JPEG compression. The error correction capability is adjusted adaptively to avoid redundancy of error correction capability and reduce the length of the error correction code.

\end{enumerate}

The remainder of this paper is organized as follows. Section \ref{section:Related Works} provides an overview of the embedding domain representation and the dither modulation algorithm. Section \ref{section:Proposed Method} describes the proposed robust image steganography. Section \ref{section:Experiment Results and Discussion} describes the experimental results, and Section \ref{section:Conclusion} concludes the paper.

\section{Related Works}
\label{section:Related Works}
In this section, we introduce the notations of steganography, the representative symbol of the embedding domain, the dither modulation method, and GMAS \citep*{yuRobustAdaptiveSteganography2020} algorithm.

\subsection{Notations}
Vectors and matrices are shown in bold in this paper. 
A cover and stego image of size $n_{1} \times n_{2}$ are represented by $\boldsymbol{X}=\left(x_{i j}\right)^{n_{1} \times n_{2}}, \boldsymbol{Y}=\left(y_{i j}\right)^{n_{1} \times n_{2}}$ correspondingly, where $i,j$ is the row and column index of the matrix.
JPEG images are used for the cover and stego images.
The image $\boldsymbol{X}$ is decompressed to the spatial domain and is represented by the symbol $J^{-1}(X)$.

\subsection{Embedding Domain Representation}

\begin{figure}[htb]
\centering
\includegraphics[width=2.5in]{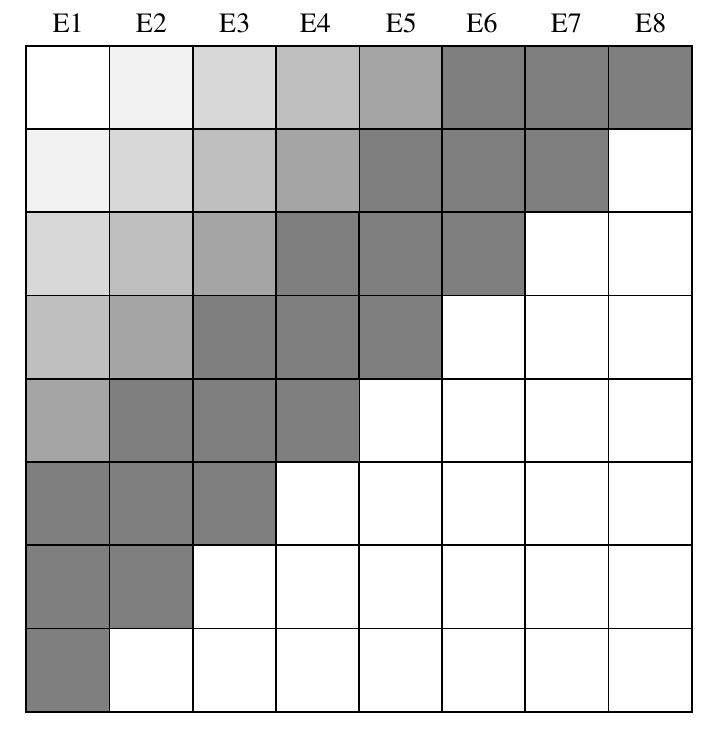}
\caption{ Embedding domain representation and symbol in an $8 \times 8$ DCT block.}
\label{fig:related:domain}
\end{figure}

In a block of $8\times8$ DCT coefficients, we indicate different embedding domains with various symbols. 
In Figure \ref{fig:related:domain}, E8 represents the eight coefficients of the counter-diagonal.
Enumber1-number2 is used to indicate the relevant embedding domain.
E6-8, for example, is a combination of E6, E7, and E8.

\subsection{Dither Modulation Algorithm}

\begin{figure}[htb]
\centering
\includegraphics[width=0.95\textwidth]{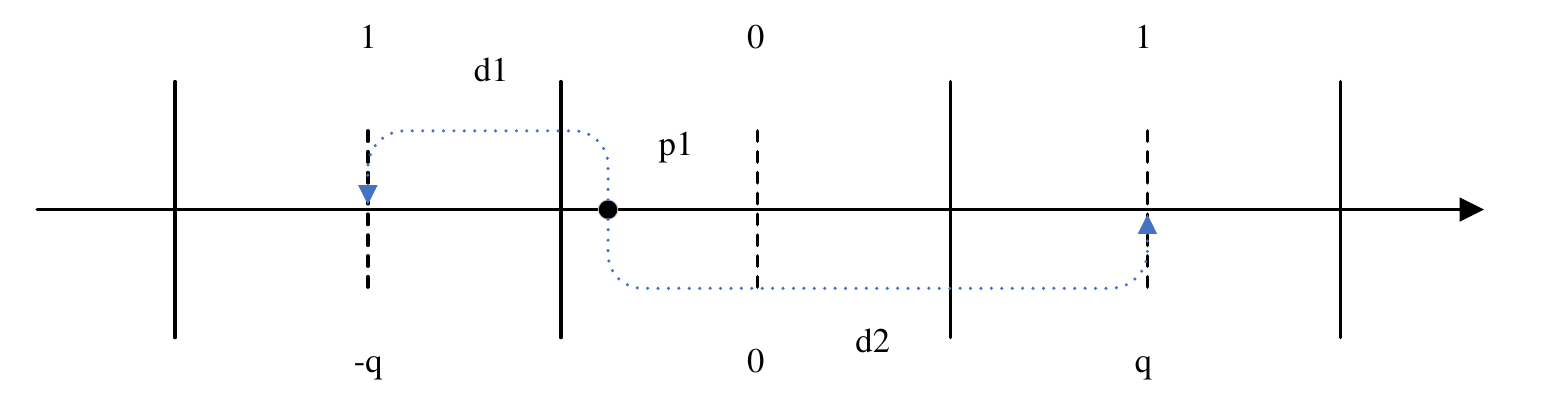}
\caption{Embedding scheme for the dither modulation algorithm.}
\label{fig:related:QIM}
\end{figure}

Dither modulation is a watermarking method that is based on Quantization Index Modulation (QIM) \citep*{chenQuantizationIndexModulation2001,nodaHighperformanceJPEGSteganography2006a}. 
As illustrated in Figure \ref{fig:related:QIM}, each de-quantized DCT coefficient is divided into several odd and even intervals based on the quantization step in its quantization table.
The de-quantized DCT coefficient is in the odd interval, indicating that the cover element is ``1" and the even interval is ``0".
The parity of quantization coefficients is used to embed message bits.
The de-quantized coefficient is adjusted using the minimal modification distance.
$p1$ is in the even interval and represents information ``0", as illustrated in Figure \ref{fig:related:QIM}. The modified distance $d1$ must be subtracted from the de-quantized coefficient if information ``1" is to be embedded.
The de-quantized coefficient adjustment distance in GMAS \citep*{yuRobustAdaptiveSteganography2020} can, however, be in two directions.

\subsection{Review of GMAS}
\label{section:Related:GMAS}
The following is a concise summary of the core concept of GMAS \citep*{yuRobustAdaptiveSteganography2020}.

\begin{enumerate}[1)]
\item 
Calculate dequantized DCT coefficients for a cover image $\boldsymbol{X}$.
\item 
Calculate the distortion $\rho_{ij}$ of all cover elements with the distortion function J-UNIWARD \citep*{holubUniversalDistortionFunction2014}, where $i,j$ is the row and column index of the matrix.

\item 
Calculate the asymmetric distortion $\rho^{+}$, $\rho^{-}$ using
\begin{equation}
\label{eq:gmas:rho}
\begin{aligned}
&\rho_{i j}^{+}= \begin{cases}\alpha \cdot \rho_{i j}, & x_{i j}<\frac{\overline{x_{i j}}}{q_{i j}} \\
\rho_{i j}, & x_{i j} \geq \frac{\frac{x_{i j}}{q_{i j}}}{\bar{x}_{i j}}\end{cases} \\
&\rho_{i j}^{-}= \begin{cases}\alpha \cdot \rho_{i j}, & x_{i j}>\frac{\overline{x_{i j}}}{q_{i j}} \\
\rho_{i j}, & x_{i j} \leq \frac{x_{i j}}{q_{i j}}\end{cases}
\end{aligned},
\end{equation}
where
$\rho_{ij}$ denotes the distortion of the $ij$th quantized DCT coefficient, $q_{ij}$ denotes the quantization step and
$\overline{x_{i j}}$ represents the de-quantized DCT coefficients, $\alpha \in[0,1]$.

\item 
With the generalized dither modulation algorithm, extract a cover sequence $C$ and the modification distances $d^{+}$, $d^{-}$ from E6-8 of all $8 \times 8$ DCT blocks.

\item 
Calculate the modifying costs $\xi^{+}$, $\xi^{-}$,
according to
\begin{equation}
\label{eq:gmas:zeta}
\begin{aligned}
\zeta_{i j}^{+} &=\frac{\rho_{i j}^{+}}{q_{i j}}, \xi_{i j}^{+}=\zeta_{i j}^{+} \times d_{i j}^{+} \\
\zeta_{i j}^{-} &=\frac{\rho_{i j}^{-}}{q_{i j}}, \xi_{i j}^{-}=\zeta_{i j}^{-} \times d_{i j}^{-}
\end{aligned},
\end{equation}
where $\zeta_{i j}$ indicates the distortion of the $ij$th de-quantized DCT coefficient.

\item 
Encode message $m$ with RS codes to get an encoded message $m^{\prime}$.
\item 
To make a stego sequence $S$, embed the encoded message $m^{\prime}$ into the cover sequence $C$ using ternary STC. Quantize de-quantized DCT coefficients of the cover image using the stego sequence $S$ and modification distances $d^{+}$, $d^{-}$. The quantized DCT coefficients can be used to generate a stego image $Y$.

\end{enumerate}

To derive the quantized DCT coefficient from the received steganographic image, the receiver employs the same quantization table as the embedding method. The encrypted data is then extracted via STC decoding. Finally, the accurate secret information is obtained via RS decoding.

\section{Proposed Method}
\label{section:Proposed Method}
In this section, we improve the robustness of JPEG steganography (adaptive embedding domain selection, adaptive error correction). The pseudocode of the proposed scheme is provided in this section.

\subsection{ Framework of the proposed scheme}

\begin{figure*}[!t]
\centering
\includegraphics[width=1\textwidth]{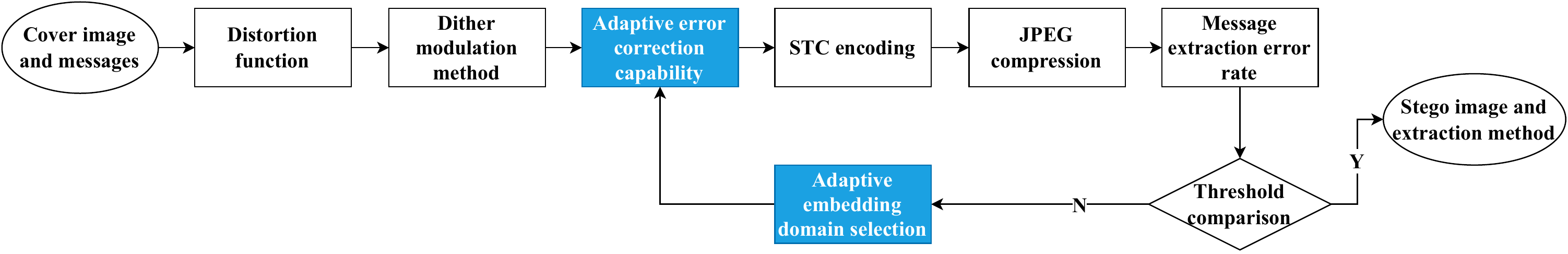}
\caption{ The framework of the proposed scheme.}
\label{fig:proposed:lct}
\end{figure*}

The framework of the proposed scheme is in Figure \ref{fig:proposed:lct}.
The distortion function calculates the modification cost of the cover image. Combined with the dither modulation algorithm, the modification distance of the embedded messages is calculated. The secret messages are encoded with small error correction capability, and the low frequency region is selected in the embedding domain. Steganographic code STC is used to produce steganographic images. Next, the quality factor of the cover or channel is used to simulate JPEG compression of the steganographic image. The secret messages are extracted from the compressed steganographic image. If the accuracy of the extracted messages reaches the set threshold, the steganographic image and the extraction method are output. If the accuracy rate of the extracted messages does not meet the set threshold, the error correction capability is given priority to increasing. If the error correction capability can not effectively improve the accuracy of the extracted messages, the embedding domain is adaptively adjusted to the high-frequency region. The error correction capability and embedding domain are adjusted iteratively until the accuracy of the extracted messages reaches the robustness threshold.

\subsection{Adaptive embedding domain selection}

\begin{figure}[htb]
\centering
\includegraphics[width=3.5in]{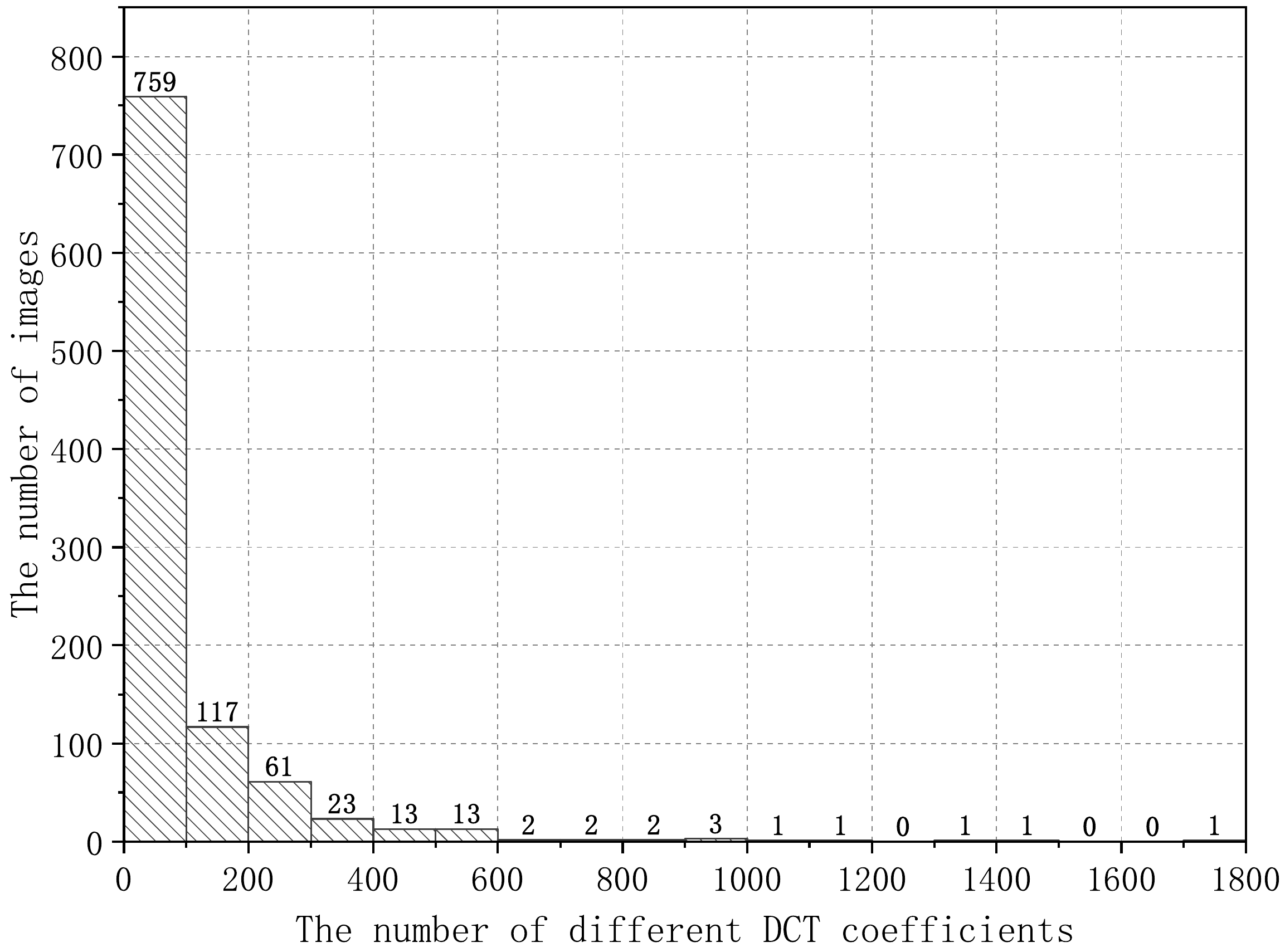}
\caption{
The robustness of images against JPEG compression. Select 1000 images at random with $QF = 75$ (denoted as $O_{75}$) from BOSSbase 1.01 \citep*{basBreakOurSteganographic2011} and obtain their compressed version (denoted as $C_{75}$) with $QF = 75$. On the horizontal axis, the number of different DCT coefficients of image $O_{75}$ and image $C_{75}$, and on the vertical axis, the number of images whose coefficients change in this interval.}
\label{fig:proposed:butong_lbx}
\end{figure}

The fixed embedding domain in the middle frequency region can not make good use of the robustness of the image itself.
For each cover image in the embedding process of GMAS \citep*{yuRobustAdaptiveSteganography2020}, a fixed embedding domain (middle embedding domain E6-8) is employed.
Under JPEG compression, different images suffer different degrees of loss and have varying levels of the robustness of JPEG compression. 
In Figure \ref{fig:proposed:butong_lbx}, most images demonstrate high robustness and little change in DCT coefficients after compression.
The low-frequency embedding domain has high anti-steganalysis performance but limited robustness, as GMAS demonstrates. The high-frequency embedding domain has a high level of robustness but a weak level of anti-steganalysis performance. The embedding domain E6-8 sacrifices anti-steganalysis performance for some robust cover images. The overall anti-steganalysis performance of GMAS is weak since the embedding domain is not modified according to the robustness of the image. The method proposed in this work changes the embedding domain selection adaptively based on the JPEG compression robustness of different images.

The embedding domain has a great impact on anti-steganalysis performance. A suitable embedding domain can improve steganography security.
We designed six embedding domains, as indicated in Table \ref{table:method:domain}, to take advantage of the robustness of various images. E64 represents the DCT coefficients for the entire block sized $8\times8$. 
$E_n$ is the number of the embedding domain, from low frequency to high frequency.
To ensure steganography security, the image with high robustness employs a low-frequency embedding domain. The high-frequency embedding domain is employed for the image with weak robustness. To balance robustness and anti-steganalysis performance, the embedding domain is gradually adjusted from low frequency to high frequency based on image robustness. This can increase the security of steganography without losing its robustness.

\begin{table}[htb]
\caption{ The represent of the embedding domain and the number of AC coefficients that can be embedded.}
\label{table:method:domain}
\centering
\begin{tabular}{c c c}
\hline
\bfseries $E_n$  & \bfseries Represent & \bfseries Number of AC coefficients \\
\hline
1 & E64 & 64 \\
2 & E2-8 & 35 \\
3 & E3-8 & 33 \\
4 & E4-8 & 30 \\
5 & E5-8 & 26 \\
6 & E6-8 & 21 \\
\hline
\end{tabular}
\end{table}

\subsection{Adaptive error correction capability}

\begin{figure}[htb]
\centering
\includegraphics[width=3.5in]{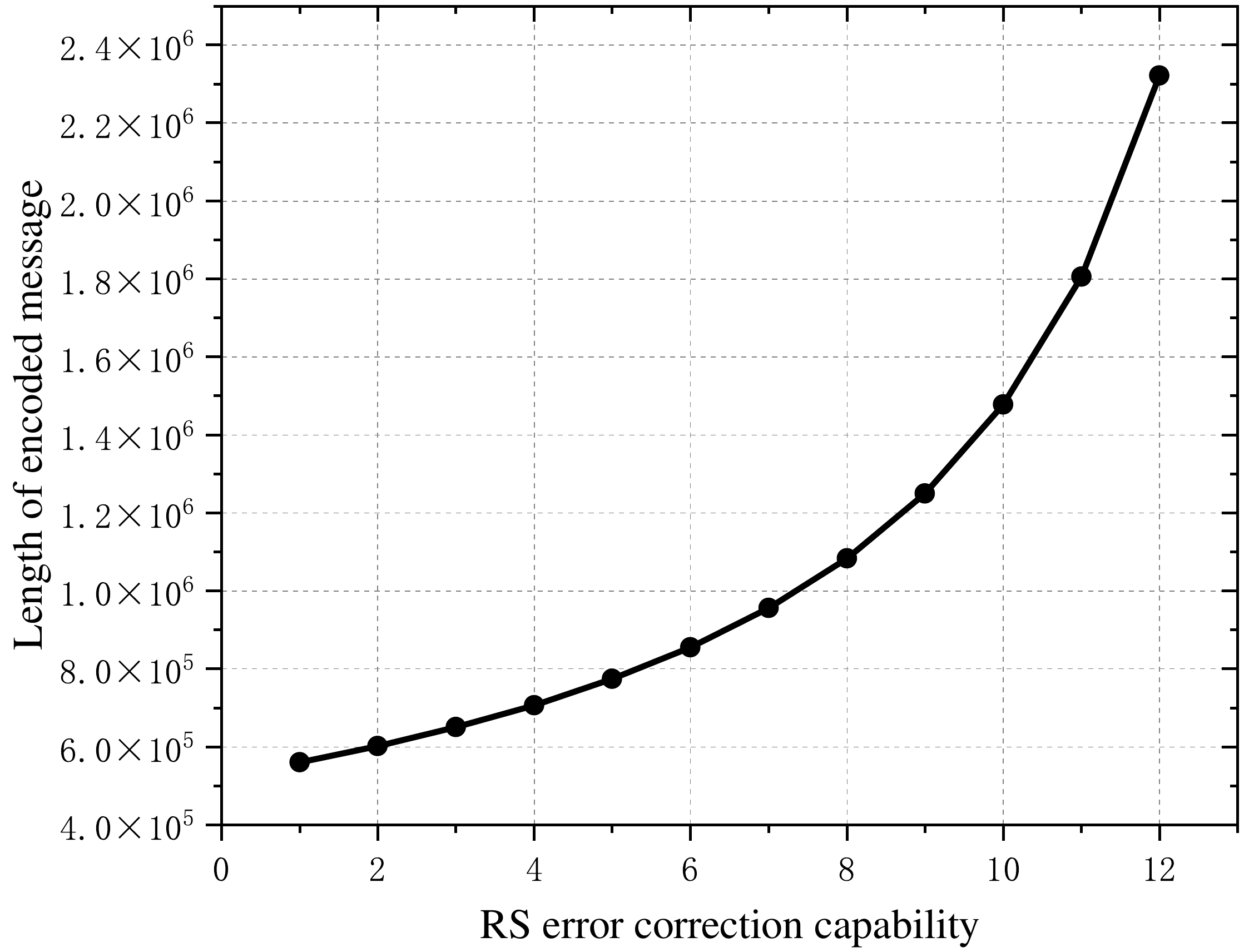}
\caption{ With varying RS code parameters, the actual length of encoded secret messages. A Secret message with a length of 524,288 ($256 \times 256 \times 8$) bits is generated at random and encoded with various error correction capabilities. The vertical axis represents the length of secret message after actual coding, while the horizontal axis represents different error correction capabilities.}
\label{fig:proposed:length_rs}
\end{figure}

Fixed error correction capability causes error correction coding redundancy and affect steganography security for images with strong robustness. For images with weak robustness, the effect of the error correction code is not obvious.
Fixed error correction capability of secret information is used in GMAS (RS \citep*{macwilliams1977theory} error correction code with error correction capability of 8).

We conduct RS coding for 524,288 bits of secret message and test the length of the encoded message.
The experimental results are shown in Figure \ref{fig:proposed:length_rs}.
The increase of error correction capability of error correction code will increase the length of encoded information. 
Because there are more embedded modifications in the image, if the error correcting capacity is utilized overly, it may not improve the robustness of steganography but it will damage the security of steganography.
If the use of error correction code is small, the error correction capability cannot be fully utilized to improve robustness. Therefore, the adaptive error correction capability proposed in this paper is very valuable for balancing robustness and anti-steganalysis performance.

The method proposed in this paper adaptively adjusts the error correction capabilities of the error correction code based on the image robustness to JPEG compression.
Error correction codes are pretty helpful in improving the robustness of steganography methods against JPEG compression.
This work employs 12 kinds of RS code error correction capabilities. 
The image with high robustness has a smaller error correction capability, which decreases the length of the error correction code, and the amount of modification to the cover image. 
The image with weak robustness fully utilizes the capabilities of the error correcting code. This can improve robustness while sacrificing little steganography security.

\subsection{Pseudo Code of the Proposed Scheme}
We give pseudo-code to help in understanding the proposed method. The following is a description of the embedding scheme:

\begin{enumerate}[1)]
\item
Calculate the original symmetric costs using distortion functions. To acquire the asymmetric costs $\rho^{+}$ and $\rho^{-}$, using the method given in Section \ref{section:Related:GMAS}.

\item
Get the modification distances $d^{+}$, $d^{-}$, and then calculate the modifying costs $\xi^{+}$ and $\xi^{-}$ using the approach described in Section \ref{section:Related:GMAS}.

\item
Initialization parameters, embedding domain using $E_n = 1$ in Table \ref{table:method:domain}, error correction capability using $t = 1$.
\item
From the embedding domain, extract cover elements $S_c$ and their related asymmetric costs $\rho^{+}$ and $\rho^{-}$.
The capability $t$ is used to encrypt the secret message $m$ and obtain the encoded message $m^{\prime}$.
To make a stego sequence $S$, embed the encoded message $m^{\prime}$ into the cover sequence $S_c$ using ternary STC.
With the stego sequence $S$ and modification distances $d^{+}$ and $d^{-}$, quantize the de-quantized DCT coefficients of the cover image. The quantized DCT coefficients can be used to create a stego image $Y$.

\item
The cover image is compressed using the quality of the channel factor $Q_{channel}$. If the channel quality factor is unknown, the image is compressed using the quality factor of the cover $Q_{cover}$.
Calculate the secret message extraction error rate $R_e$ after extracting the secret message.
End the loop and execute step 7) if $R_e$ meets the robustness threshold $T_r$.
If $R_e$ fails to meet the robustness threshold $T_r$, set the error correction capability to $t + 1$ and go back to the previous step 4).
If the error correction capability $t$ reaches 12, go to the next step 6).

\item
The embedding domain adjusts $E_n + 1$, initializes the error correction capability $t = 1$, and goes back to the previous step 4).
Execute the next step 7) when $E_n = 6$.

\item
Output the steganographic image, embedding domain $E_n$, and error correction capability $t$.
\end{enumerate}

The receiver receives the steganographic image $Y$ compressed by the channel quality factor $Q_{channel}$.
The extracting scheme is described as follows:
\begin{enumerate}[1)]
\item
Calculate the de-quantized DCT coefficients of the stego image $Y$. Dither modulation algorithm with embedding domain $E_n$ is used to extract the stego sequence $S_s$.
\item
STC decoding is used to recover the secret message encoded by RS from $S_s$.
\item
To extract secret message $m$, RS coding with $t$ error correction capability is employed.

\end{enumerate}

\section{Experiment Results and Discussion}
\label{section:Experiment Results and Discussion}

In this section, we first introduce the experimental settings in Section \ref{section:Experiment:Setups}. The results and comparisons with previous methods in terms of robustness, anti-steganalysis performance and social network application are presented in Section \ref{section:Experiment:robustness} to Section \ref{exp:Social network application}.
The robustness threshold parameter experiment is shown in  Section \ref{section:Determining the robustness threshold}.
Ablation experiments are shown in Section \ref{section:Performances of adaptive embedding domain}  and Section \ref{section:Performances of adaptive error correction capability}.

\subsection{Setups}
\label{section:Experiment:Setups}

\begin{table}[htbp]
\caption{ Experimental settings.}
\label{table:exp:settings}
\centering
\resizebox{\textwidth}{!}{
\begin{tabular}{ll}
\hline
Image source           & Bossbase 1.01 
 \\
Image size             & $512 \times 512 $                          \\
Image type             & Grayscale                           \\
                                   
Message                & Randomly generated binary sequences \\
RS parameters          &  $n^{*} = 31$                                   \\
Distortion Function & J-UNIWARD                           \\
Detection feature      & CCPEV, DCTR   
\\
Robustness threshold $T_r$ & 0.0001  
\\
 Compared algorithms & 

\makecell[c]{
GMAS\citep*{yuRobustAdaptiveSteganography2020}
\\E-DMAS\citep*{zhangEnhancingReliabilityEfficiency2020}
\\ MR-DMAS\citep*{yinRobustAdaptiveSteganography2021}
\\SS\citep*{zhuRobustSteganographyModifying2019}\\
ESS\citep*{qiaoRobustSteganographyResisting2021}
\\SSR\citep*{wuSignSteganographyRevisited2022}
}

\\
\hline
\end{tabular}
}
\end{table}

Bossbase 1.01 \citep*{basBreakOurSteganographic2011} was the image data set used in this experiment. There are 10000 grayscale images in the collection, each sized $512 \times 512$ pixels.
Convert the images to JPEG images.
The embedded payload is $p = n_m/n_{nzac}$, where $n_m$ is the length of secret information and $n_{nzac}$ is the number of non-zero AC coefficients.
The extraction error rate is $R_{error} = n_{error}/n_m$, where $n_{error}$ is the number of error information bits.
The JPEG compression quality factor of the channel is $Q_{channel}$, whereas the quality factor of cover is $Q_{cover}$.
The steganalysis feature set includes CCPEV \citep*{kodovskyCalibrationRevisited2009} and DCTR \citep*{holubLowComplexityFeaturesJPEG2015}.
Selection of STC secure parameter $h = 10$.

The FLD ensemble \citep*{kodovskyEnsembleClassifiersSteganalysis2012a} is used to train the detectors as binary classifiers with default settings. The ensemble minimizes the overall classification error probability $P_{\mathrm{E}}=\min _{P_{\mathrm{FA}}} \frac{1}{2}\left(P_{\mathrm{FA}}+P_{\mathrm{MD}}\right)$, where $P_{\mathrm{FA}}$ and $P_{\mathrm{MD}}$ are the false-alarm and missed-detection probabilities, respectively, when all priors are equal. The average error rate $\bar{P}_{\mathrm{E}}$ calculated across ten $2000/2000$ database splits determines the ultimate level of anti-steganalysis performance, with a larger $\bar{P}_{\mathrm{E}}$ signifying stronger anti-steganalysis performance. Table \ref{table:exp:settings} shows the experimental setup in detail.

\subsection{Comparison of robustness}
\label{section:Experiment:robustness}

\begin{figure*}[htbp]
\centering
\subfigure[$Q_{channel} = 85$]{
    \label{fig:exp:lbx_Q85} 
    \includegraphics[width=0.45\textwidth]{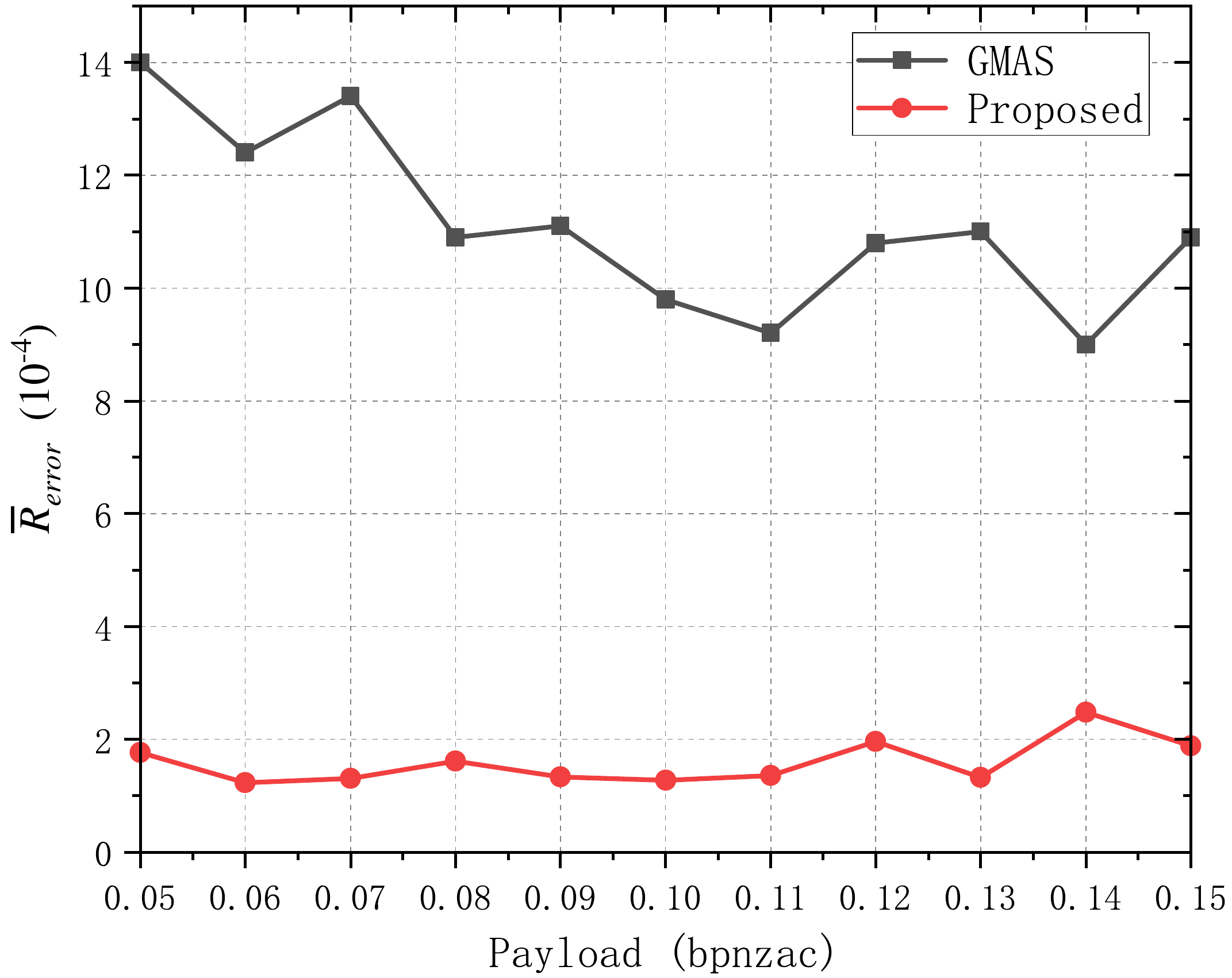}}
\subfigure[$Q_{channel} = 95$]{
    \label{fig:exp:lbx_Q95} 
\includegraphics[width=0.45\textwidth]{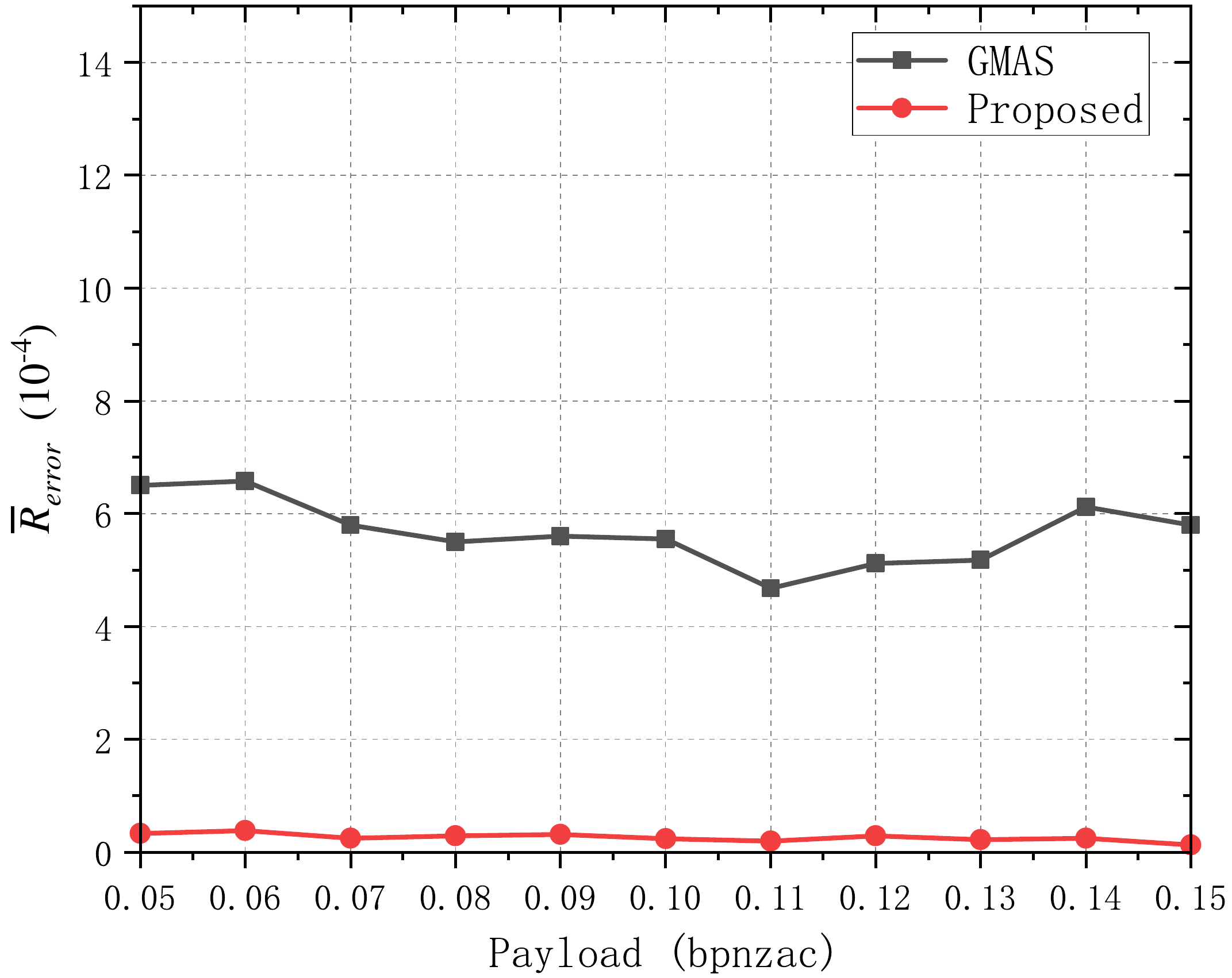}}
\caption{ When $Q_{cover} = 65$, $Q_{channel} = 85$ and $Q_{channel} = 95$, and the payload is 0.05 to 0.15, 1000 images are randomly selected from BOSSbase 1.01.
Average extraction error rates of the proposed scheme and GMAS \citep*{yuRobustAdaptiveSteganography2020} with (a) $Q_{channel} = 85$ and (b) $Q_{channel} = 95$ were calculated.}
\label{fig:exp:lbx_Q85_Q95} 
\end{figure*}

The method proposed in this paper is based on GMAS. Firstly, the robustness of JPEG compression with different quality factors is compared.
When $Q_{cover} = 65$, $Q_{channel} = 85$ and $Q_{channel} = 95$, and the payload is 0.05 to 0.15, 1000 images are randomly selected from BOSSbase 1.01. Average extraction error rates of the proposed scheme and GMAS \citep*{yuRobustAdaptiveSteganography2020} were calculated.
In Figure \ref{fig:exp:lbx_Q85_Q95}, the proposed scheme outperforms GMAS. Compared to GMAS, the proposed method is much more robust under a variety of payloads. Because the method takes use of the inherent robustness of the image.

\begin{figure}[htb]
\centering
\includegraphics[width=3.5in]{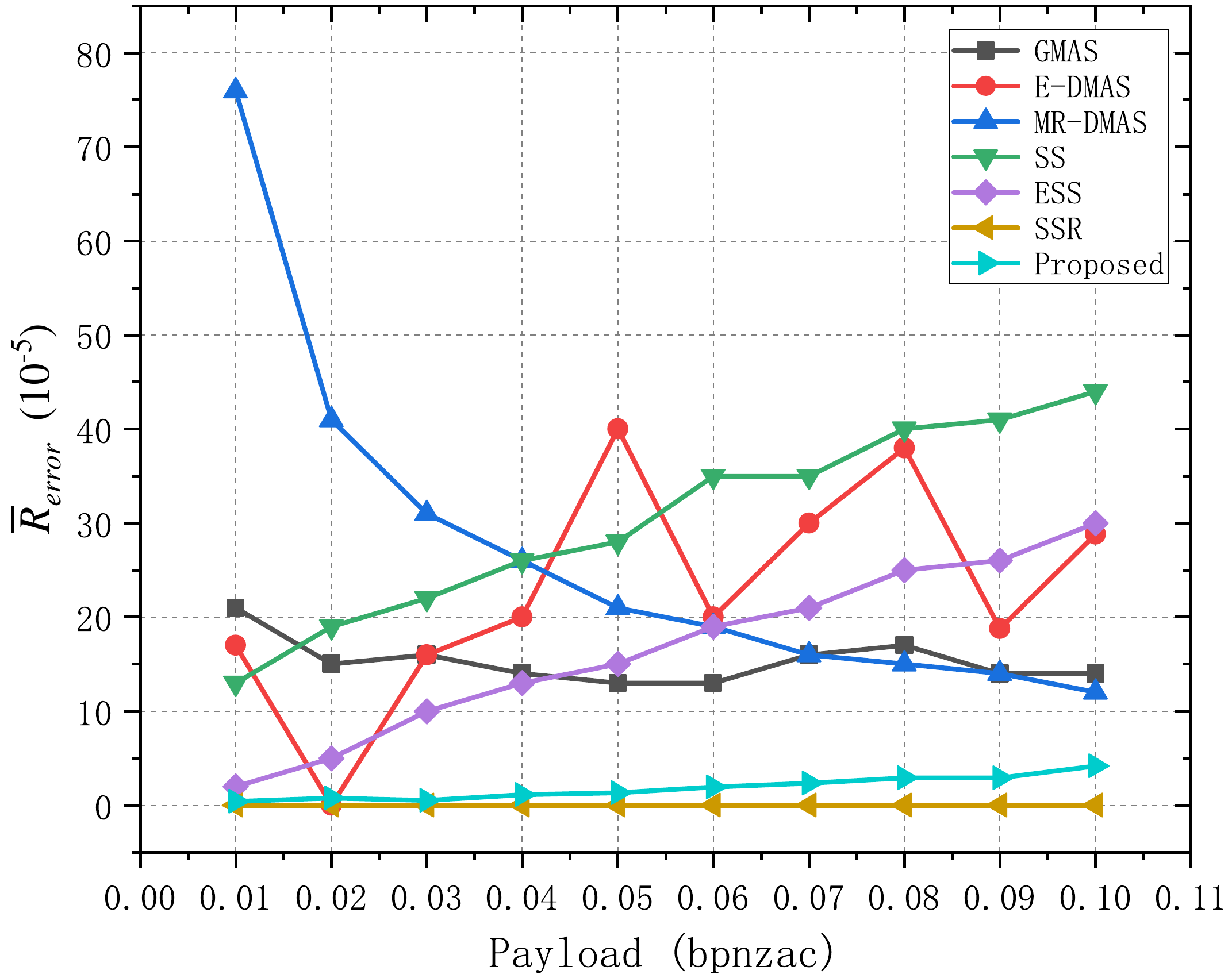}
\caption{When $Q_{cover} = 75$, $Q_{channel} = 75$, and the payload is 0.01 to 0.10, 1000 images are selected from BOSSbase 1.01.
Compare the average error rate with the recent robust steganography method.}
\label{fig:exp:lbx_3}
\end{figure}

Then, we compare the error rate of message extraction with the recent ``Cover robustness" steganography method.
When $Q_{cover} = 75$, $Q_{channel} = 75$, and the payload is 0.01 to 0.10, 1000 images are selected from BOSSbase 1.01.
Compare the average error rate with the recent robust steganography method.
The average secret message extraction error rates are shown in Figure \ref{fig:exp:lbx_3}. 

The robustness of the proposed method is better than other methods and slightly lower than SSR \citep*{wuSignSteganographyRevisited2022}.
The robustness of the proposed method is high, mainly because the algorithm takes advantage of the robustness of the image itself against JPEG compression, and uses the embedding domain and error correction capability suitable for this image.
As the payload increases, the error rate of message extraction increases steadily.
SSR \citep*{wuSignSteganographyRevisited2022} makes use of the constant sign coefficient before and after JPEG compression. The sign of AC coefficient is difficult to change in the compression process, and it is very robust.
GMAS \citep*{yuRobustAdaptiveSteganography2020} encrypts secret messages using ternary STC and an error correction algorithm. As the payload grows, the possibilities of the number of mistakes in the decoded secret messages outnumber the error correction capabilities. 
The E-DMAS \citep*{zhangEnhancingReliabilityEfficiency2020} embedding process uses the error correcting code twice. When the error correcting code is flawed, robustness suffers significantly.
In MR-DMAS \citep*{yinRobustAdaptiveSteganography2021}, with the increase of payload, the instability coefficient of re-compression adjustment increases, which 
reduces the loss of the cover caused by compression and greatly reduces the error rate of extracted information.
In SS \citep*{zhuRobustSteganographyModifying2019}, as the payload increases, the probability of the coefficient from zero to non-zero increases, resulting in a gradual decrease in robustness.
Because ESS \citep*{qiaoRobustSteganographyResisting2021} is a channel-dependent version of SS, it performs better when $Q_{channel}$ is available.

\subsection{Comparison of anti-steganalysis}
\label{section:Experiment:security}

\begin{figure*}[htbp]
    \centering
  \subfigure[CCPEV]{
    \label{fig:exp:ccpev_3} 
    \includegraphics[width=0.45\textwidth]{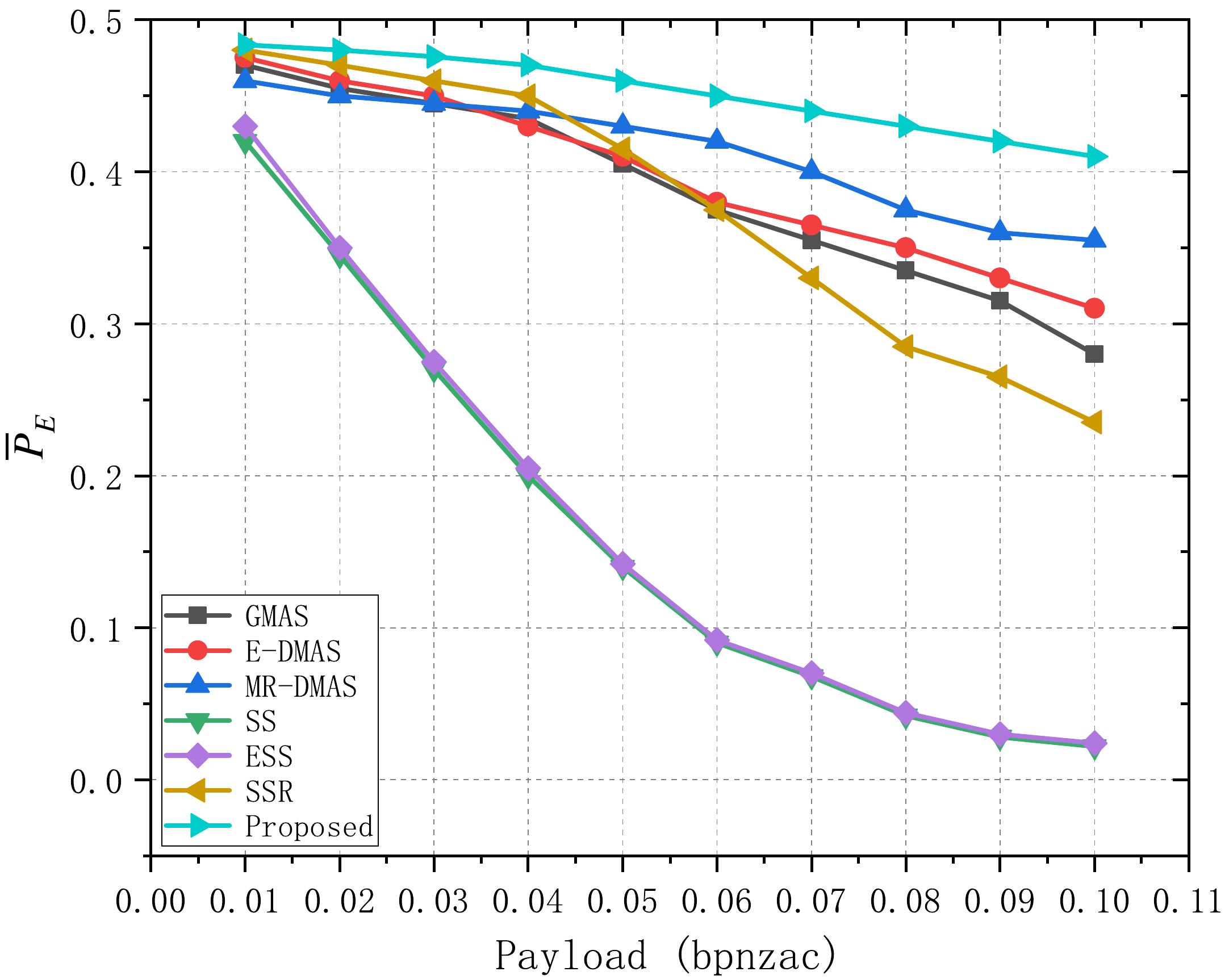}}
  \subfigure[DCTR]{
    \label{fig:exp:dctr_3} 
    \includegraphics[width=0.45\textwidth]{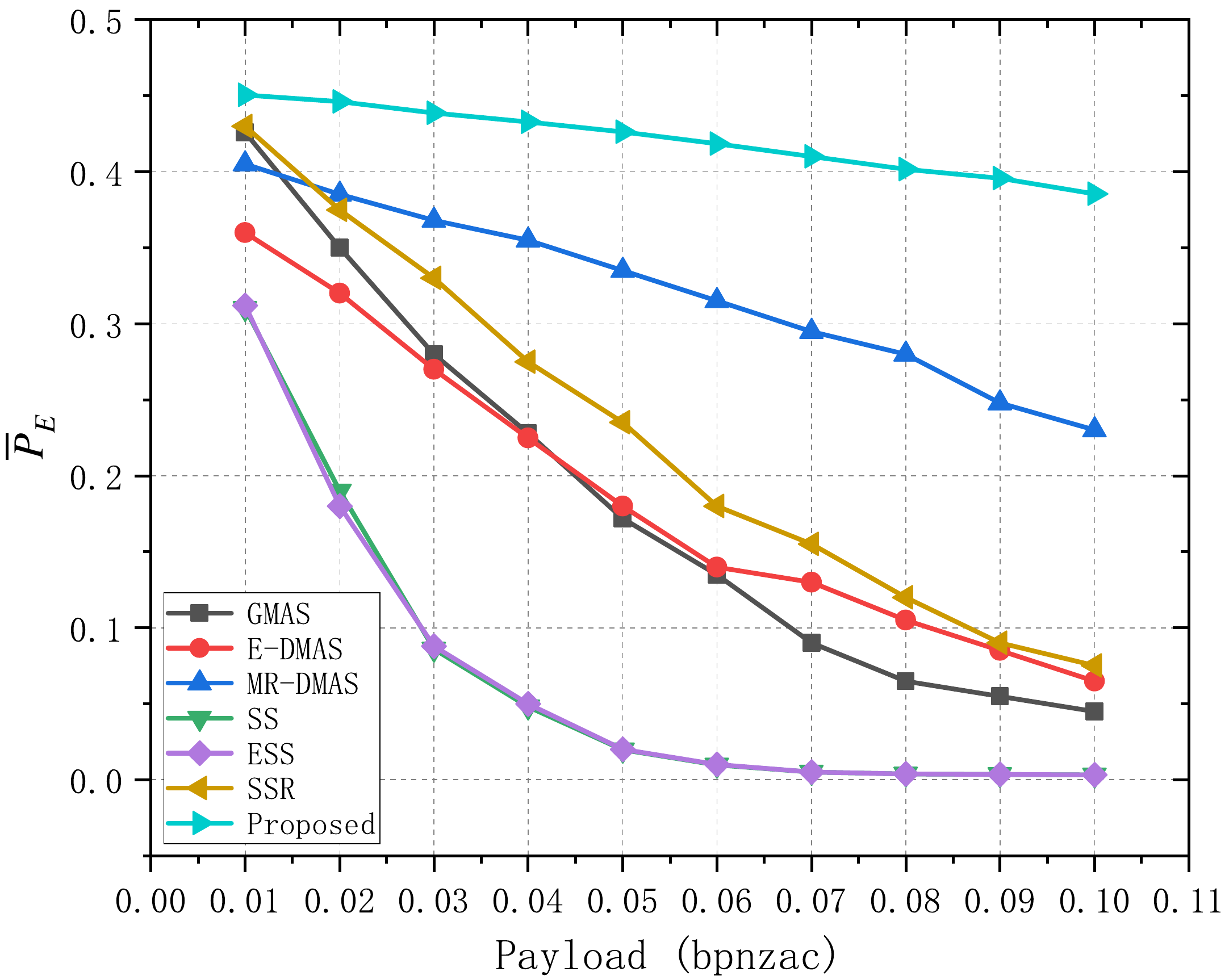}}
  \caption{When $Q_{cover} = 75$, $Q_{channel} = 75$, and the payload is 0.01 to 0.10, 2000 images are selected from BOSSbase 1.01.
Compare the average detection error rates $\bar{P}_{\mathrm{E}}$ with the recent robust steganography method.
}
  \label{fig:exp:aqx_ccpev_dctr_3} 
\end{figure*}

When $Q_{cover} = 75$, $Q_{channel} = 75$, and the payload is 0.01 to 0.10, 2000 images are selected from BOSSbase 1.01.
Compare the average detection error rates $\bar{P}_{\mathrm{E}}$ with the recent robust steganography method.
Figure \ref{fig:exp:aqx_ccpev_dctr_3} exhibit the anti-steganalysis of different algorithms when using CCPEV and DCTR, respectively.

The anti-steganalysis of the proposed method is superior to other methods. 
The proposed method embeds information in low frequency regions as much as possible.
Because the low frequency region has high anti-steganalysis performance.
The proposed method gives priority to small error correction capability, according to image robustness.
Thus, the length of the error correction code is shortened, the change of the cover is reduced, and the anti-steganography performance is improved.
GMAS \citep*{yuRobustAdaptiveSteganography2020} performs well in steganalysis because it combines asymmetric distortion and ternary STC.
The E-DMAS \citep*{zhangEnhancingReliabilityEfficiency2020} embedded domain is at high frequency, and the error correction code length is large, and the0 anti-steganography is seriously degraded.
In MR-DMAS \citep*{yinRobustAdaptiveSteganography2021}, the re-compression adjustment coefficient has an extensive modification range, although the embedding domain of MR-DMAS is the low frequency region with excellent anti-steganalysis performance. 
The anti-steganalysis performance reduces rapidly as the payload increases.
In SS \citep*{zhuRobustSteganographyModifying2019},ESS \citep*{qiaoRobustSteganographyResisting2021} and SSR \citep*{wuSignSteganographyRevisited2022}, the embedding method is to invert the coefficient symbol, which can greatly modify the image and has low anti-steganalysis performance.

\subsection{Social network application}
\label{exp:Social network application}

We investigate the robustness of our proposed scheme using one of the most widely used social network platforms, Facebook.
The constantly evolving JPEG encoder used by Facebook will resize and recompress submitted images based on their sizes and quality parameters. If the image's $Q_{cover}$ is no larger than 85 and it is less than $512 \times 512$, Facebook will recompress it with $Q_{channel} = 71$; otherwise, $Q_{channel} = 71$ varies from image to image.
We submit 60 images to Facebook (selected at random from BOSSbase 1.01 with $Q_{cover} = 60, 65, 70$, each containing 20 images).
Table \ref{table:exp:FaceBook} displays the results, where $N_{success}$ denotes the number of images from which the messages can be completely retrieved. As expected, our scheme is robust and can be effectively utilized in the real world.

\begin{table}[htb]
\caption{
The proposed method and GMAS in terms of $\bar{R}_{\text {error }}$ and $N_{success}$ versus Facebook with 60 images randomly picked from BOSSbase 1.01 with $Q_{cover} = 60, 65, 70$ and each containing 20 images at 0.1 bpnzac.
}
\label{table:exp:FaceBook}
\centering
\begin{tabular}{ccccc}
\hline
& Quality Factor & 60 & 65 & 70 \\
\hline
\multirow{2}{*}{proposed}  & $\bar{R}_{\text {error }}$  & 0  & 0  & 0  \\
                      & $N_{success}$  & 20 & 20 & 20 \\
\multirow{2}{*}{GMAS} &  $\bar{R}_{\text {error }}$  &  0.0024  & 0.0012   &  0.0001  \\
                      &  $N_{success}$  &  16  &  15  &  19 \\
\hline
\end{tabular}
\end{table}

\subsection{Determining the robustness threshold}
\label{section:Determining the robustness threshold}

\begin{figure}[htb]
\centering
\includegraphics[width=3.5in]{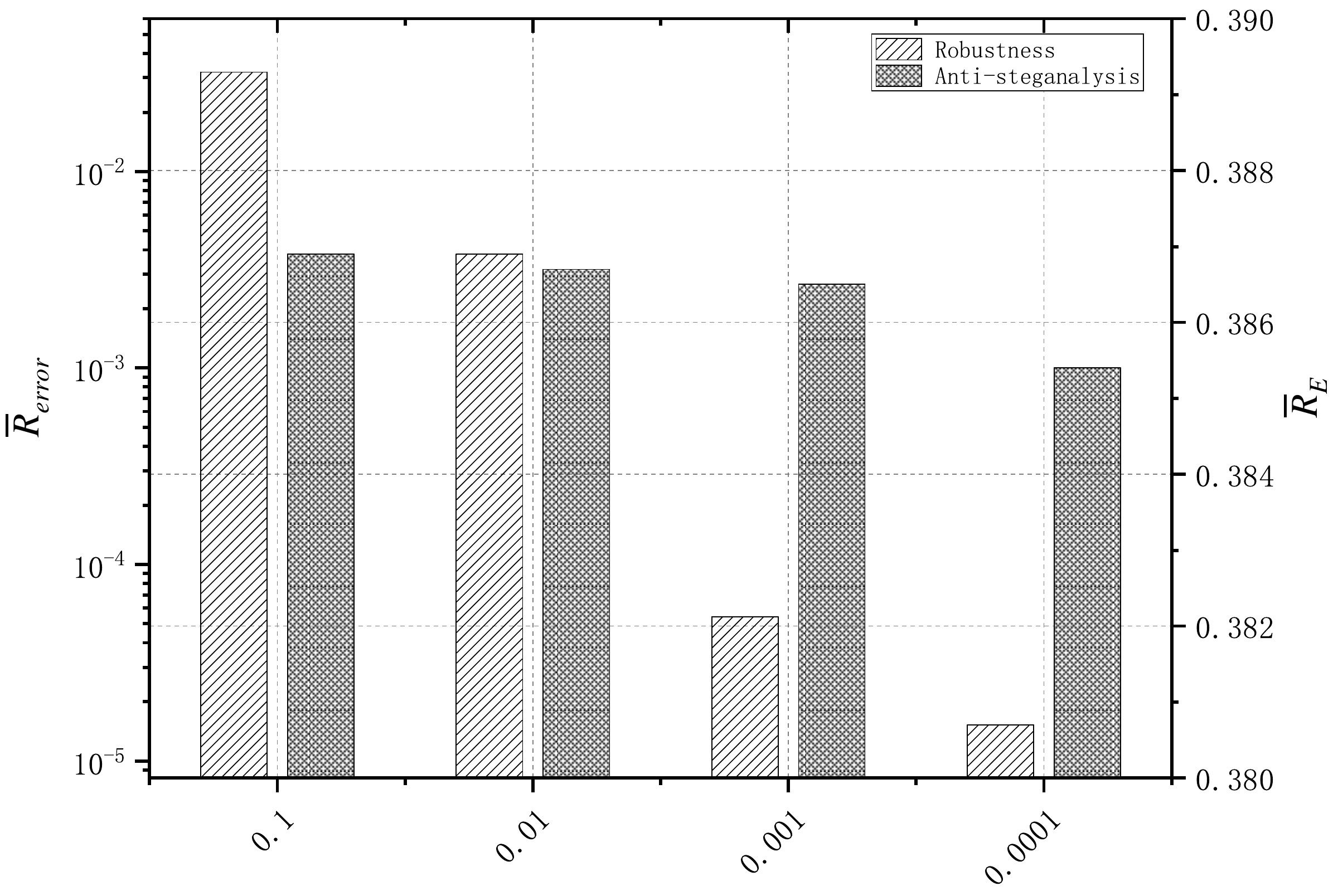}
\caption{ The robustness and anti-steganalysis performance of different robustness thresholds.}
\label{fig:exp:lbx_yuzhi}
\end{figure}

In order to select an appropriate robustness threshold $T_r$, 2000 images are randomly selected from BOSSbase 1.01 \citep*{basBreakOurSteganographic2011}, with $Q_{cover} = 75$.
The message embedding payload is 0.1.
To detect steganographic images, we utilize the feature DCTR \citep*{holubLowComplexityFeaturesJPEG2015} and set $Q_{channel} = 75$ to simulate channel JPEG compression.
As shown in Figure \ref{fig:exp:lbx_yuzhi}, with the decrease of threshold, the robustness is improved significantly. 
The proposed algorithm iterates to find the best embedding domain and error correction capability as the threshold gets lower, ensuring that the extraction error rate satisfies the threshold requirement.
As the threshold decreases, the anti-steganalysis decreases little. Because the method in this paper is to ensure anti-steganalysis first, and then to improve robustness. The priority of embedding is low frequency embedding domain and small error correction capability.
We select $T_r = 0.0001$ as the threshold of the algorithm.

\subsection{Ablation test on adaptive embedding domain selection}
\label{section:Performances of adaptive embedding domain}
\begin{figure}[htb]
\centering
\includegraphics[width=3.5in]{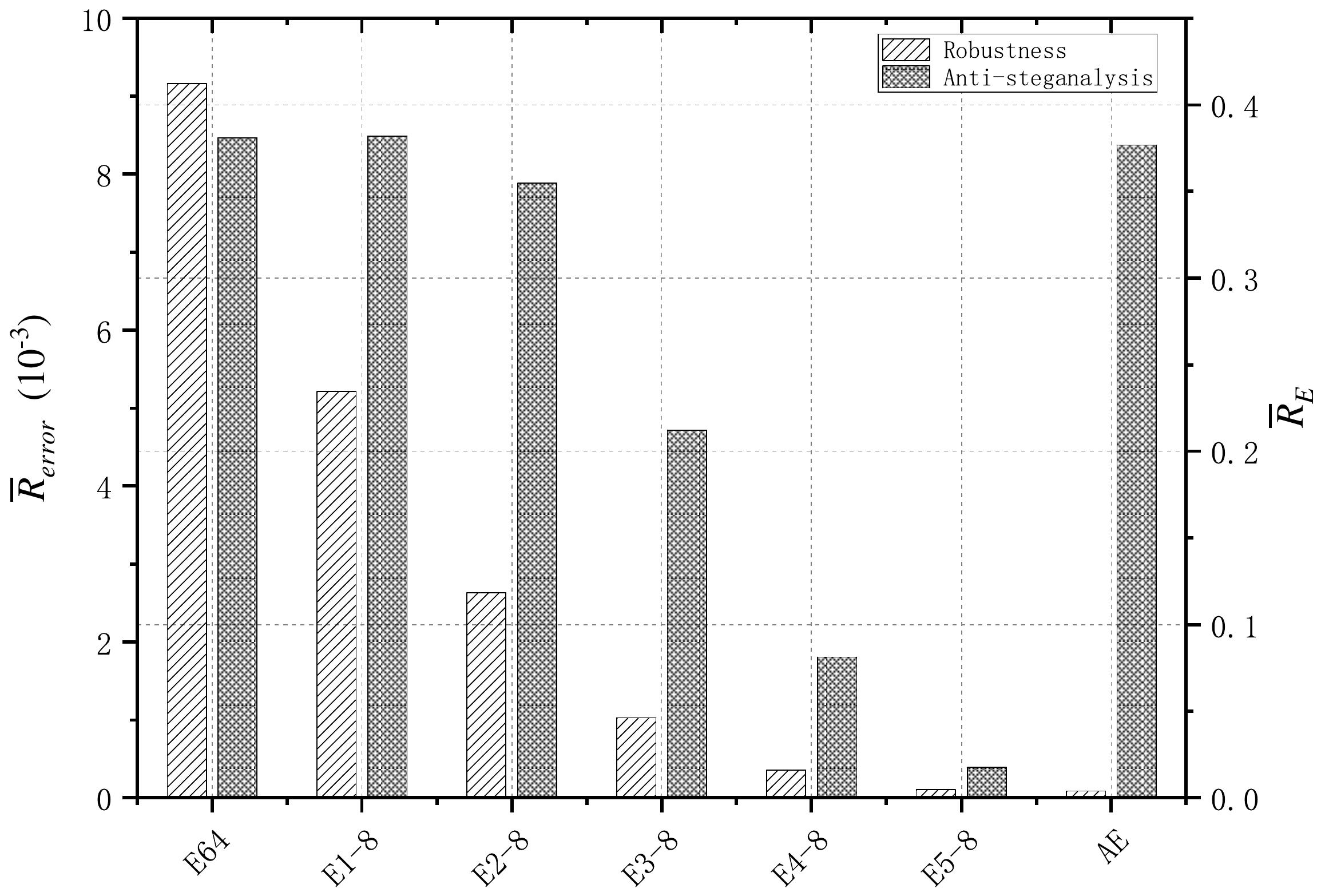}
\caption{ The robustness and anti-steganalysis performance of different embedding domains.}
\label{fig:exp:qianyuyu_xiaoguo}
\end{figure}

To verify the performance of the improved adaptive embedding domain approach, 2000 images are randomly selected from BOSSbase 1.01 \citep*{basBreakOurSteganographic2011}, with $Q_{cover} = 75$.
The message embedding payload is 0.15 and the RS error correction capability is fixed at 8.
To detect steganographic images, we utilize the feature DCTR \citep*{holubLowComplexityFeaturesJPEG2015} and set $Q_{channel} = 75$ to simulate channel JPEG compression. As shown in Figure \ref{fig:exp:qianyuyu_xiaoguo}, we gradually increase the embedding domain from low frequency to high frequency, where AE stands for adaptive embedding domain selection adjustment method.

The robustness increases when the embedding domain changes from low to high frequency, but the anti-steganalysis performance loss is severe. The increase in robustness is not apparent when the embedding domain is in the low frequency region. The robustness of the embedding domain improves significantly when the frequency is high. However, when the embedding domain moves from low to high frequency, the anti-steganalysis performance of the method rapidly degrades.
As a result, the embedding region must be adjusted adaptively based on the robustness of the image.
The low-frequency embedding domain is utilized for images with high robustness, whereas the high-frequency embedding domain is used for images with weak robustness. The embedding domain is modified based on the image robustness, which not only assures robustness but also ensures that anti-steganalysis performance is not sacrificed, achieving a balance of robustness and anti-steganalysis performance.

\subsection{Ablation test on adaptive error correction}
\label{section:Performances of adaptive error correction capability}

\begin{figure}[htb]
\centering
\includegraphics[width=3.5in]{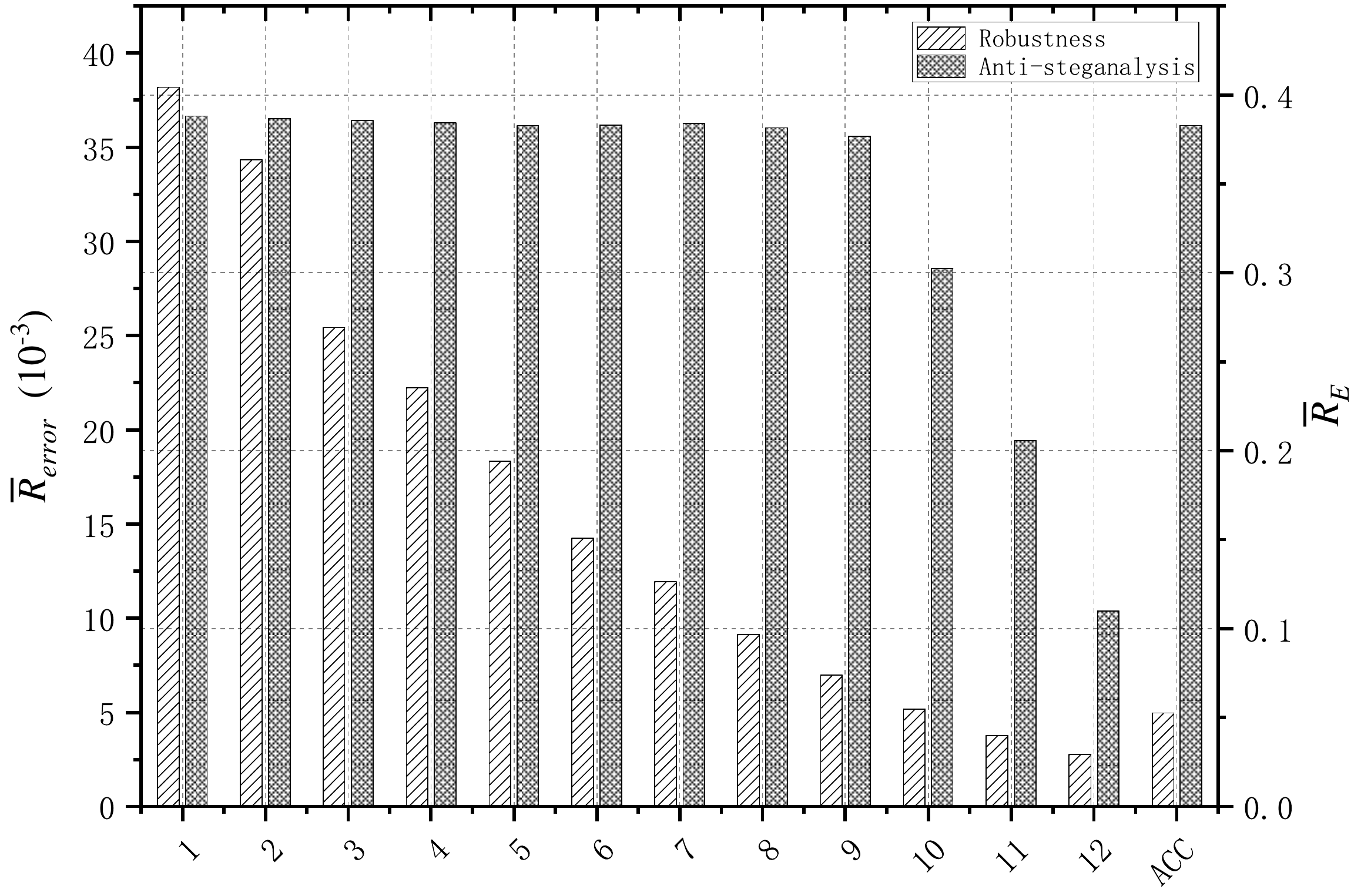}
\caption{ The robustness and anti-steganalysis performance of different error correction capabilities.}
\label{fig:exp:jiucuoma_xiaoguo}
\end{figure}

To verify the performance of the improved adaptive error correction approach, 2000 images are randomly selected from BOSSbase 1.01 \citep*{basBreakOurSteganographic2011}, with $Q_{cover} = 75$.
The message embedding payload is 0.15 and the embedding domain is set to E64.
To detect steganographic images, we employ the feature DCTR \citep*{holubLowComplexityFeaturesJPEG2015} and set $Q_{channel} = 75$ to simulate channel JPEG compression. The horizontal ordinate is the error correction capability, and ACC is the adaptive error correction capability adjustment scheme, as illustrated in Figure \ref{fig:exp:jiucuoma_xiaoguo}.

The robustness of the proposed scheme is gradually improving as the error correction capacity rises from low to high. In the case of low error correction capability, the loss of anti-steganalysis performance is small.
When the error correction capability exceeds 9, anti-steganalysis performance rapidly degrades.
ACC is an adaptive error correction scheme that guarantees not only high anti-steganalysis performance but also high robustness. As a result, fixing the embedding domain first and adaptively adjusting the error correction capability may make greater use of the robustness of the image. When the robustness of the image is high, a poor error correction capability can be used to achieve the desired effect. There is no need for excessively extensive error correction coding, which expands the image modification extent while sacrificing anti-steganalysis performance.

\section{Conclusion}
\label{section:Conclusion}

Based on the robustness of the image, we propose a JPEG compression-resistant steganography method in this paper.  We adaptively adjust the embedding domain of the dither modulation algorithm to balance the robustness and anti-steganalysis performance of the method. Furthermore, we adaptively adjust the capability of the error correction code, lower the length of redundancy in the error correction code, and increase robustness without compromising too much anti-steganalysis performance. The proposed method outperforms GMAS \citep*{yuRobustAdaptiveSteganography2020} in terms of robustness and anti-steganalysis performance, according to the results of the experiments.

We plan to expand our effort to color images and increase the embedding capabilities in the future. We will also explore the connection between image robustness and JPEG compression.

\section*{Acknowledgements}
\label{sec::Acknowledgments}
\par This work was supported in part by the Natural Science Foundation of China (Grant 62172001, Grant 61860206004) and the Shenzhen R\&D Program (Grant JCYJ20200109105008228).

\bibliography{template}

\end{document}